\documentclass{aa}
\usepackage{graphicx}

\newcommand{\Mv}    {M$_{\rm V}$}  
\newcommand{\mv}    {{\rm M_{V}}}  
\newcommand{\BV}    {(B--V)}  
\newcommand{\bv}    {{\rm (B-V)}}  
\newcommand{\Msun}  {M$_\odot$}  
\newcommand{\zsun}  {{\rm Z}_\odot}  
\newcommand{\post}  {{\rm post}}  
\newcommand{\prior} {{\rm prior}}  

\begin{document}

   \title{An inverse method to interpret colour-magnitude diagrams}
   \titlerunning{Inversion of colour-magnitude diagrams.}

   \author{J.-L. Vergely
          \inst{1}
          \and
          J. K\"oppen\inst{1,2,3}
          \and
          D. Egret\inst{1}
          \and
          O. Bienaym\'e\inst{1}
          }

   \offprints{D. Egret}
   \mail{Daniel.Egret@astro.u-strasbg.fr}

   \institute{Observatoire Astronomique de Strasbourg, 
              UMR 7550, Universit\'e Louis Pasteur, Strasbourg, France 
          \and
              Institut f\"ur Theoretische Physik und Astrophysik
              der Universit\"at, D-24098 Kiel, Germany 
          \and
              International Space University, Parc d'Innovation,
              F-67400 Illkirch, France
             }

   \date{Received 13 November 2000 / Accepted 20 February 2002}

   \abstract{An inverse method is developed to determine the star formation
      history, the age-metallicity relation, and the IMF slope from
      a colour-magnitude diagram.
      The method is applied to the Hipparcos HR diagram. We found 
      that the thin disk of our Galaxy shows 
      a peak of stellar formation 1.6 Gyr ago. The stars close to the Sun  
      have a solar metallicity and a mean IMF index equal to 
      3.2. However, the model and the evolutionary tracks do not 
      correctly reproduce the horizontal giant branch. 
     }

    \maketitle
  

\section{Introduction}

  The Hertzsprung-Russell diagram or its directly-observable
  counterpart, the colour-magnitude diagram (CMD) contains information
  about the age distribution of a stellar population, the initial mass
  function (IMF), and the metallicity of the stars. The distribution
  of the stars in such a diagram is essential for our understanding
  of the formation and evolution of the galaxy to which the stars
  belong.

  In order to infer the star formation history (SFH), one usually
  tries to match the CMD as closely as possible by model populations
  computed with various scenarios for the SFH. The quality of the
  agreement between observed and predicted CMDs is assessed either
  qualitatively by visual inspection or quantitatively, e.g. by the
  $\chi^2$-test (cf. Dolphin, 1997).
  Often, one represents the SFH by a number of discrete values or
  by a suitable analytic form, and adopts either a fixed metallicity 
  or a certain age-metallicity relation. This approach, which could be
  called the direct or synthetic method, has been widely used 
  (Tolstoy et al. 1993, Gallart et al. 1996, Haywood et al. 1997).
  Since one usually keeps the number of free parameters for the 
  underlying model as small as necessary or practical, the adopted 
  functional forms for the SFH and age-metallicity relation (AMR) 
  impose certain limitations on 
  what type of model populations are possible to be considered. Thus 
  the obtained best fit is only optimal in the context of the adopted 
  model, which may not necessarily represent the real galaxy.
  In what sense the interpretation will be limited by these
  constraints is difficult to assess accurately, and can only be
  done by experimentation.
  In most recent work (Harris \& Zaritsky 2001, Dolphin 2001)
  these limitations are being overcome by using more efficient 
  minimization techniques.
  The technique of inverse methods (Craig \& Brown, 1986; Tarantola 
  \& Valette, 1982a,b; Twomey, 1977) deals with such a problem in the 
  opposite way: One tries to determine the functional form of e.g. 
  the SFH with as much freedom as possible -- with a resolution that 
  depends on the information contained in the data --  under the 
  constraint that it matches the observed data. Such a method 
  has been applied to the CMD by Hernandez et al. (1999): this 
  work presents a non-parametric method for the 
  maximum likelihood solution of the SFH, through the iterative 
  solution of an integro-differential equation. 
  The method presented here is similar and has the advantage 
  of determining not only the SFH from CMD but also the IMF 
  slope and the AMR with a temporal resolution as high as possible. 
  Specific tools are presented in order to estimate 
  the validity of the inverse procedure like the a posteriori 
  covariance and the resolution. We apply our method to the Hipparcos data 
  for solar neighbourhood stars in order to give constraints on the Galaxy 
  thin disk formation.

\section{Relation between SFH and the HR diagram}
  \label{s:relation}

  The distribution of stars in a CMD is determined by the 
  following ingredients:
  \begin{itemize}
    \item the SFH $\psi(t)$ specifies the number of stars of all 
          masses born as a function of age $t$;
    \item the initial mass function (IMF) gives the number $N$ of stars
          in each generation per unit interval in stellar
          mass $M$. A convenient form is a power-law
          $$
            dN \,\,\, \propto \,\,\, M^{-\Gamma} \,\, dM
          $$
          where the index $\Gamma = 2.35$ refers to Salpeter's (1955)
          function. As usual, it is assumed to be independent of time;
    \item due to the chemical evolution of a galaxy, the chemical
          composition of the gas from which stars are born changes with
          time. This is described by an age-metal\-ici\-ty relation (AMR)
          $Z(t)$;
    \item the results of stellar evolution calculations give the
          properties -- temperature, luminosity --
          of a star of given mass and metallicity for any given time
          after its birth;
    \item stellar atmosphere models give the photometric colours
          for any star.
  \end{itemize}

  For comparison with models, one bins the CMD in some suitable way in
  colour index and magnitude. Then the number density of stars in 
  a given bin $q$ can be written as 
  \begin{equation}
      \label{e:stars}
         D_q  = \int \psi(t) F_q(t, Z(t), \Gamma ) dt   
  \end{equation}
  the superposition of the contributions from all stellar 
  populations of single age $t$, metallicity $Z$, and IMF slope 
  $\Gamma$. Since the densities $F_q(t, Z, \Gamma)$ would be computed 
  from stellar evolutionary tracks and the photometric colours,
  the interpretation of the observations $D_q = D_q({\rm obs})$
  consists in solving Eqn. \ref{e:stars} for the SFH $\psi$.

  In principle, one would like to determine all three functions SFH, 
  IMF, and AMR, i.e. functions defined on an infinite number of points. 
  However huge the observed set of data might be, it remains finite. 
  Thus, even if there were no noise and observational uncertainties, 
  one can derive the three functions only up to a certain limit of detail.

  The usual approach is to assume that the functions have some specified
  analytical form described by a few parameters -- such as the exponent
  in the power-law for the IMF, or a simple SFH with a few rectangular
  starbursts -- and the objective is simply to find the best model of 
  this type that reproduces the observations. In doing so, one does not 
  know of what (time) resolution could be reached with the data available,
  and thus one may not make full use of what the observations could furnish.

\section{The inverse method}

  The determination of the full functional form of the SFH
  (or AMR or IMF) from a CMD is an under-determined problem,
  because the amount of observational data is only finite and thus
  cannot provide the information for every detail of the function.
  Application of a straight inversion technique to Eqn. \ref{e:stars}
  could be very sensitive to the noise in the data, and could well
  give mathematically correct but unphysical results 
  (Craig \& Brown 1986). The problem can be regularized by demanding 
  that the true solution is smooth in some suitably quantified sense 
  (Twomey 1977, Craig \& Brown 1986). Thus one reduces the number of 
  free parameters, and the problem becomes well-posed. 

  Tarantola \& Valette (1982a) use a Bayesian approach which describes 
  how the a priori knowledge about the functions -- which might 
  be the null information -- is changed by the information contained 
  in the observational data. If the data contain no information about 
  a parameter, one would merely retrieve its a priori value.
  The a posteriori probability density $f_{\post}(M \vert D)$ for the 
  vector $M$ containing the unknown model parameters, given the observed 
  data $D$, is linked by Bayes' theorem
  $$
       f_{\post}(M \vert D) \propto L(D\vert M) \,\cdot \, f_{\prior}(M)
  $$
  to the likelihood function $L$ and the prior density function
  for the parameter vector. The factor of proportionality is
  obtained by normalization $\int f_{\post}(M \vert D) dM = 1$
  over all parameter space.

  The theoretical model shall be described by an operator $g$ which
  connects the model parameters $M$ with the predicted data
  $$
       D_{\rm predicted} = g(M)
  $$
  which is to agree as closely as possible with the observed data.
  If we assume that both the prior probability and the errors in the
  data are distributed as Gaussian functions, the posterior
  distribution becomes
  \begin{eqnarray}
      f_{\post}(M\vert D)
        & \propto & \exp(-{1\over 2}(D-g(M))^{T}C_D^{-1}(D-g(M))
                                     \nonumber \\
        & & \ \ \ \ \ \ \ \
            -\frac{1}{2}(M-M(0))^{T}C_0^{-1}(M-M(0)) )  \nonumber
  \end{eqnarray}
  The prior is specified by the value $M(0)$ of the parameter 
  vector and its variance-covariance matrix $C_0$.
  The matrix $C_D$ specifies how the variances in the
  observed data are obtained (e.g. if a non-Gaussian 
  distribution of the observational errors is to be
  approximated) and whether one needs to take into
  account correlations between the individual data. 

  In this method the model parameters may include single value 
  parameters as well as entire functions. For simple parameters, 
  $C_0$ is the matrix of its a priori variances and a priori 
  covariances between each parameter. 
  The variances and covariances would be chosen to be large, if 
  we do not have any initial knowledge about the parameter values. 
  On the other hand, if the values are accurately known, one would 
  use variances correspondingly smaller. 
  For a function, $C_0$ is a functional operator which has for its
  kernel the auto-correlation function. Most commonly one uses
  a Gaussian kernel:
  $$
     C_0(x,x')  =  \sigma_{0}(x)\sigma_{0}(x')
                \exp\left(-\frac{(x-x')^{2}}{\xi_{0}^2}\right)
  $$
  $\sigma_0(x)$ -- which can be any function of $x$ -- describes
  how strongly the solution is allowed to fluctuate between points,
  and $\xi_0$ is the smoothing length for the solution. 
  Other types of kernels may be employed, e.g. an exponential kernel 
  which would smooth the solution differently.

  The best estimator $\tilde{M}$ for $M$ is the most probable value
  of $M$, knowing the set of data $D$. This condition is reached by
  minimizing the quantity
  \begin{eqnarray} 
      \frac{1}{2}(D-g(M))^{T}C_D^{-1}(D-g(M)) & & \cr
      +\frac{1}{2}(M-M(0))^{T}C_0^{-1}(M-M(0)) & & 
  \end{eqnarray}

  which corresponds to a maximum likelihood condition.\\

  Since the operator $g$ is non-linear, the solution for the parameter
  vector must be done iteratively (Tarantola \& Valette 1982a, 1982b):
  \begin{eqnarray}
     \label{e:valette}
      \tilde{M}(k+1) & = & M(0)  +  C_0 G^{T}(k)
                (C_D + G(k) C_0 G^{T}(k))^{-1}     \nonumber \\
                 & &  (D + G(k)[\tilde{M}(k)-M(0)]-g(\tilde{M}(k)))
  \end{eqnarray}
  with $k$ counting the number of iterations, and the matrix of
  partial derivatives
  $$
      G(k)=\frac{\partial g(M(k))}{\partial M}
  $$
  For use in subsequent equations we shall abbreviate:
  $$
      S := C_D + G(k)\cdot C_0\cdot G^{T}(k)
  $$

\subsection{Formulation of the problem}

  The CMD is binned both in \Mv\ and \BV, by dividing the range
  of interest of \Mv--values into $m$ intervals of equal length
  $\Delta \mv$. Likewise, the range of \BV\ values is divided into
  $n$ intervals of equal length $\Delta$\BV.
  Then, $D(i,j)$ is the number of stars having their
  magnitude \Mv\ between  $\mv_i$ and $\mv_i + \Delta\mv$
  as well as their colour index \BV\ between \BV$_j$ and
  \BV$_j$ + $\Delta$\BV. For a less cumbersome notation, we use a 
  single index $q = (i-1)n+j = 1 \dots nm$ corresponding to 
  bin $(i,j)$.

  Synthetically-generated stellar populations for single ages $t$, 
  metallicities $Z$, and IMF slopes $\Gamma$ are used to specify 
  which fraction 
  of the total number of stars is found in each bin $q$
  \begin{equation}
     \label{e:density}
           F_q(t, Z, \Gamma )\, \Delta\mv\, \Delta\mbox{\BV}
  \end{equation}
  Convolving these populations with the SFH by Eqn. \ref{e:stars}
  gives the number density $D_s$ of stars in that bin.

  Since the star formation rate is always positive, changing over
  from $\psi$ to $\alpha(t)$
  \begin{equation}
  \label{par1}
      \alpha(t) := \ln(\psi(t)/\psi_0)
  \end{equation}
  where $\psi_0$ is a constant, will enforce this physical fact.
  We shall assume that the uncertainties in the function $\alpha(t)$
  follow a Gaussian law. This implies that the errors in $\psi(t)$
  follow a log-normal distribution.

  Since the position of isochrones in the CMD is a nearly logarithmic
  function of the age, it is more appropriate to use $u := \lg(t)$ instead 
  of the linear age $t$. With $ dt = 10^u \,\ln(10) \,du $ and the 
  abbreviation $h(u) := 10^u\,\ln(10)$, Eqn. \ref{e:stars} is written as
  \begin{eqnarray}
        \label{e:starsu}
           D_q & =  & \int \psi_0 \exp(\alpha(u)) 
                     F_q(u,Z(u),\Gamma) h(u) du \nonumber \\
               & =: & g_{q}(M) 
  \end{eqnarray}
  This defines $g_q$ as a non-linear operator acting on the parameter
  vector $M$ which consists of the unknown functions $\alpha(t)$ and 
  $Z(t)$, and the unknown parameter $\Gamma$:
  $$
         M=\left(\begin{array}{c}
                     \alpha(t)     \\
                     Z(t)          \\
                     \Gamma        \\
                  \end{array} \right)
  $$
  We note in passing that the IMF could of course also be
  taken as an unknown function $\phi(m)$. This would merely
  increase greatly the dimension of parameter space, resulting
  for the numerical solution of the equations in greater
  memory requirements and unacceptable execution times.
  Furthermore, an arbitrary IMF would be largely degenerate
  with the SFR, at least on the main sequence.

\subsection{The base models}
   \label{s:base}

  The operator $g_q$ (Eqn. \ref{e:starsu}) generates the model CMDs from 
  a set of base models $F_q(u,Z,\Gamma)$ (Eqn. \ref{e:density}) by 
  integrating over all ages. The base models are computed from 
  synthetic stellar populations of a single age $t$, 
  metallicity $Z$, and specified IMF; they give the probability 
  of finding a star in bin $q$ in the CMD, depending on the model 
  parameters, but we also take into account observational selection 
  effects:
  \begin{itemize}
     \item synthetic theoretical stellar populations are computed
           from the isochrones of Bertelli et al. (1994) which directly
           give the positions in the (\Mv , \BV )--diagram for stars in 
           the mass range 0.6 to 120 \Msun;
     \item the IMF is a power-law between the stellar mass limits
           of 0.6 and 80  \Msun;   
     \item the stars are uniformly and randomly distributed in space;
     \item the limiting magnitude is 8.0 mag in V band.
  \end{itemize}
  A grid of models was computed, each comprising 2 million stars
  (before applying selections), for any combination among 
  50 ages distributed linearly between $u = 6.6$ and 10.3, 10 
  metallicities between  $Z = 0.0004$ and $0.05$,
  and 5 IMF slopes between 2.0 and 4.5.
  This was done through linear interpolation in age 
  and metallicity
  from the Padua isochrones.

  All integrals are evaluated numerically, with the 
  rectangle method being sufficiently accurate.

\section{Determination of SFH and IMF slope}

  We shall first consider the case of deducing simultaneously
  the SFH and the slope of the IMF: The parameter vector
  $M$ consists of the unknown function $\alpha(t)$
  and the unknown parameter $\Gamma$:
  $$
      M  = \left( \begin{array}{c}
                     \alpha(t)    \\
                     \Gamma       \\
                  \end{array}
           \right)
  $$
  The metallicity is considered to be known and constant. 

  Then, the matrix of partial derivatives is 
  $$
     G(k)=\left( \begin{array}{cc}
                   \frac{\partial g_1}{\partial \alpha}
                        &\frac{\partial g_1}{\partial \Gamma}\\
                   \frac{\partial g_2}{\partial \alpha}
                        &\frac{\partial g_2}{\partial \Gamma}\\
                   \vdots      & \vdots \\
                   \frac{\partial g_{s}}{\partial \alpha}
                        &\frac{\partial g_{s}}{\partial \Gamma} \\
                \end{array}
          \right)
  $$
  with $s = nm$ the total number of data bins.
  
  Equation \ref{e:valette} gives the procedure to improve an estimate
  of the parameters. For better legibility, we drop the dependence on
  the iteration number $k$ of $G$, $\alpha(u)$, and $\Gamma$.
  The $(i,j)$-th component of the matrix product is (for all
  $i,j = 1, \dots s$)
  \begin{eqnarray}
        (G C_0 G^{T})_{i,j} & =
        &\int \int C_{\alpha}(u,u') \psi_0^2 \exp(\alpha(u) +
             \alpha(u')) \nonumber \\
      & & \ \ \ \ \ \ \times F_i(u,\Gamma) F_j(u',\Gamma) h(u) h(u')du du'
                         \nonumber \\
      & &  + \left( \int \psi_0 \exp(\alpha(u))
                {\partial F_i(u,\Gamma) \over \partial \Gamma}
                  h(u)du \right)
                         \nonumber \\
      & & \times   \sigma^2_{\Gamma}
               \left( \int \psi_0 \exp(\alpha(u))
               {\partial F_j(u,\Gamma) \over \partial \Gamma}h(u) du \right)
                   \nonumber
  \end{eqnarray}
  The variance-covariance matrix $C_0$ of the prior distribution
  of the parameters is
  $$
    C_0=\left(\begin{array}{cc}
                C_{\alpha}   & 0                  \\
                0            & \sigma^2_{\Gamma}  \\
              \end{array}
        \right)
  $$
  and the covariance function $C_{\alpha}$ 
  \begin{equation}
     \label{e:calpha}
       C_{\alpha}(u,u')=\sigma^2_{\alpha}
           \exp\left(-\frac{(u-u')^2}{\xi_{\alpha}^2}\right)
  \end{equation}
  with $\xi_{\alpha}$ the correlation length in $\lg(t)$.

  The $i$-th component of the vector $V(k) = D-g(M)+G(k)(M-M(0))$, which
  depends on the iteration index $k$, is
  computed from
  \begin{eqnarray}
        V_i =  D_i
            &&  + \int \psi_0 \exp(\alpha(u))(\alpha(u)-1)
                             F_i(u,\Gamma) h(u)du   \nonumber \\
            &&  + \int \psi_0 \exp(\alpha(u))
                        (\Gamma-\Gamma_0)
                        \frac{\partial F_i(u,\Gamma)}{\partial \Gamma}
                        h(u)du 
                        \nonumber
  \end{eqnarray}
  Defining the vector $W(k) = S^{-1} V(k)$, the $k+1$-st estimate for the
  parameters is computed from their $k$-th values by
  \begin{eqnarray} 
      \alpha_{k+1}(u) =\sum_i W_i(k)
        && \int C_{\alpha}(u,u') \psi_0 \exp(\alpha_k(u')) \nonumber \\
        && \times  F_i(u',\Gamma_k)h(u')du' \nonumber
  \end{eqnarray} 
  and
  \begin{eqnarray} 
       \Gamma_{k+1} = \Gamma_0 + \sum_i W_i(k) \sigma_{\Gamma}^2
       && \int \psi_0 \exp(\alpha_k(u')) \nonumber \\
       && \times   \frac{\partial F_i(u',\Gamma_k)}{\partial \Gamma}
                          h(u')du'  \nonumber 
  \end{eqnarray} 
  The estimator for $\psi$ from $\alpha$ is given by the relation
  (e.g. Saporta, 1990):
  $$
       \tilde{\psi}(u)=\psi_0 \exp \left(\alpha(u)+
                      \frac{\sigma_{\alpha(u)}^2}{2} \right)
  $$
  with the dispersion of the posterior distribution for $\alpha(t)$
  $$
        \sigma_{\psi(u)}=\psi_0
              \sqrt{ \exp(2 \alpha(u)+\sigma_{\alpha(u)}^2)
              \left(\exp(\sigma_{\alpha(u)}^2)-1 \right)}
  $$
  where $\sigma_{\alpha(u)}$ is the dispersion of the posterior
  distribution of $\alpha$ (Eqn. \ref{e:sigalf} below).

  In the computations, the derivatives such as
  the matrix $G$ and $\partial F/\partial \Gamma$ are 
  evaluated numerically. The functions $\partial F/\partial \Gamma$ 
  are interpolated by a 4 th order polynomial.

  \subsection{The posterior variance-covariance matrix}

  The internal errors made by the method on the estimated parameter
  values can approximately be computed. This is done by a second-order
  expansion of the posterior density distribution in the neighbourhood
  of the best solution:
\begin{equation}
\label{postvar}
      C_{\tilde{M}} = C_0-C_0 G^{T}S^{-1} G C_0
\end{equation}
  For the simple parameter of the IMF slope one gets
  $$
      \sigma_{\tilde{\Gamma}} = \sigma_{\Gamma} \sqrt{(1-G^{T}S^{-1}G)}  ,
  $$
  and for the function $\alpha(u)$ :
  \begin{equation}
     \label{e:sigalf}
      \sigma_{\alpha(u)} = \sqrt{\sigma^2_{\alpha}-C_{\alpha}G^{T}S^{-1} 
                                   G C_{\alpha} }  .
  \end{equation}  

  \subsection{Resolving kernel and mean index}

  Observational errors, like the finite photometric accuracy, will
  degrade the information contained in the CMD, and thus cause an
  additional loss of resolution in age. The concept of the resolving 
  kernel, introduced by Backus \& Gilbert (1970), permits us to compute 
  how much a stellar population for a given age is spread out in age. 
  Suppose that we knew the true model $M_{\rm true}$, then the observed 
  data would be
  $$
      D  = g(M_{\rm true})  .
  $$
  If the operator $g$ were linear, so that $G_{k} = G = g$,
  then Eqn. \ref{e:valette} shows how the deviation of the
  true model parameters from the initial guess is expected to be
  degraded into the observed one
  \begin{equation}
      \label{e:kernel}
         \tilde{M}-M_0 = C_0 G^{*} S^{-1} G(M_{\rm true}-M_0) .
  \end{equation}
  The operator -- called the resolving kernel --  
  $$
       K(u,u') := C_0 G^{*} S^{-1} G
  $$
  describes this degradation of information. Though it 
  applies strictly for linear models only, it remains a good
  approximation in our problem as well: the density of stars
  in a bin is a linear function of the SFH. Because the
  values for the SFH remain within a factor of 10 or so, 
  the logarithmic relation with the parameter $\alpha$ 
  does not represent a strong non-linearity.

  The relation between the resolving kernel K and the a posteriori
  variance-covariance operator $C_{\tilde{M}}$ is computed as follows 
  from Eqn. \ref{postvar}:
  $$
    C_{\tilde{M}} = (I-C_0 G^{T}S^{-1} G) C_0 = (I-K) C_0
  $$
  this yields:
  $$
    K=I-C_{\tilde{M}} C_0^{-1}
  $$
  $I$ is the Dirac function if the unknown parameter is a function (the SFH
  in our case).
  So, if the posterior variance-covariance operator increases, the amplitude
  of the resolving kernel decreases and the width increases.

  Another important and useful concept is a measure of the information 
  present in the data. This is closely linked to the resolving kernel: 
  Suppose that the mean value of the SFH is within the width 
  of the kernel $K$, so Eqn. \ref{e:kernel} gives
  $$
      (\tilde{M}-M(0))(u)= (M_{\rm true}-M(0))_{\rm mean} \cdot
                          \int K(u,u') du'  .
  $$
  The integral is called the {\it mean index} $I(u)$
  $$
      I(u) := \int K(u,u') du'
  $$
  which has the following meaning: If $I(u)$ has very low values,
  the expected model would be close to the initial guess,
  irrespective of how much the initial model would differ from the
  true model. The quality of the SFH estimator thus is poor.
  But if $I(u) \approx 1$ the deduced model would be close to the
  true model.

  We note that if we take an a priori variance-covariance operator 
  with too large a smoothing length $\xi_{\alpha}$, we will obtain 
  a good mean index. However, the a posteriori resolution will be poor.

\section{Simultaneous derivation of SFH and AMR}

   We now address the problem of determining simultaneously
   two functions, the SFH and the AMR. The parameter vector is 
   $$
       M = \left(\begin{array}{c}
                    \alpha(t)       \\
                    Z(t)            \\
                 \end{array}
           \right) .
   $$
   The variance-covariance matrix of the parameter priors is taken as:
   $$
       C_0 = \left(\begin{array}{cc}
                       C_{\alpha}   & 0       \\
                       0            & C_Z     \\
                   \end{array}
             \right)
   $$
   i.e. we assume that SFH and AMR are a priori uncorrelated.
   This is done for simplicity. We note that chemical evolution of
   the galaxy would impose a strong relation between the two functions
   and could be used to compute such a correlation.

   $C_{\alpha}$ is given by Eqn. \ref{e:calpha}, and $C_Z$ is given by
   $$
        C_Z(u,u')=\sigma^2_{Z}\exp\left(-\frac{(u-u')^2}{\xi_Z^2}\right)   .
   $$
   The derivation is nearly identical as before, with $\alpha(u)$ and
   $Z(u)$ depending on the iteration index $k$:
   \begin{eqnarray}
      && (G C_0 G^{*})_{i,j} =  \int \int C_{\alpha}(u,u') \psi_0^2
                              \exp(\alpha(u)+\alpha(u'))   \nonumber \\
      && \ \ \ \times F_i(u,Z(u))F_j(u',Z(u'))
                      h(u) h(u')du du'      \nonumber \\
      && + \int \int C_{Z}(u,u') \psi_0^2
             \exp(Z(u)+Z(u')) \nonumber \\
      && \ \ \ \times  \frac{\partial F_i(u,Z(u))}{\partial Z}
                  \frac{\partial F_j(u',Z(u'))}{\partial Z} 
                        h(u) h(u') du du'  \nonumber
   \end{eqnarray}
   In addition to the iteration for the SFH which proceeds as before,
   we have for the AMR 
   \begin{eqnarray}
      Z_{k+1}(u) & = & Z_0(u) + \sum_i W_i(k)
                 \int C_{Z}(u,u') \psi_0 \exp(Z_k(u')) \nonumber \\
      && \ \ \ \ \ \ \times   F_i(u',Z_k(u')) h(u')du'   \nonumber 
   \end{eqnarray}
   where $Z_0(t)$ is the a priori age-metallicity-relation. 

\section{Validation of the method by simulations.}

  In practice, the large amount of computing time necessary makes
  it prohibitive to determine simultaneously the SFH, AMR, and the
  IMF slope. Therefore, in the following tests we show an
  inversion of SFH and IMF slope only.
  Also, to limit the number of bins and hence the computing time,
  we consider only the histogram of the \BV -- colour index. The colour
  of the turn-off is very sensitive to the age of a stellar population,
  thus one obtains a good resolution of the SFH. 
  The range from $-0.3$ to 1.7 is divided into bins of 0.02 mag width,
  the observational error. It is not useful to use cell sizes smaller 
  than this error.
  
  The variance-covariance matrix $C_D$ for the data is assumed
  to be diagonal, as we shall assume that there are no correlation
  between the numbers of stars in different bins. To approximate
  the Poisson noise in the stellar numbers, we take the
  variance equal to the number of stars in a bin.

  Since our method is an iterative procedure, it yields models 
  that approach the data successively closer with each step. 
  The distance between model prediction and observed data can 
  be measured by the value of $\chi^2$:
  $$
     \chi^2 = \sum_s {(g_s(M) - D_s)^2 \over D_s}
  $$
  The convergence is usually monotonic. 
  Typically 10 iterations suffice to achieve a value of 
  $\chi^2$ stable within $10^{-2}$ for the determination of the SFH.
  
  In our formalism, the errors in the data are assumed to be normally
  distributed. This is an approximation valid only if each bin contains
  a large number of stars, whereas in reality the errors in the histogram 
  are distributed with a Poisson law. 

\begin{figure*}
   \begin{center}
     \includegraphics[angle=0,totalheight=6.0in]{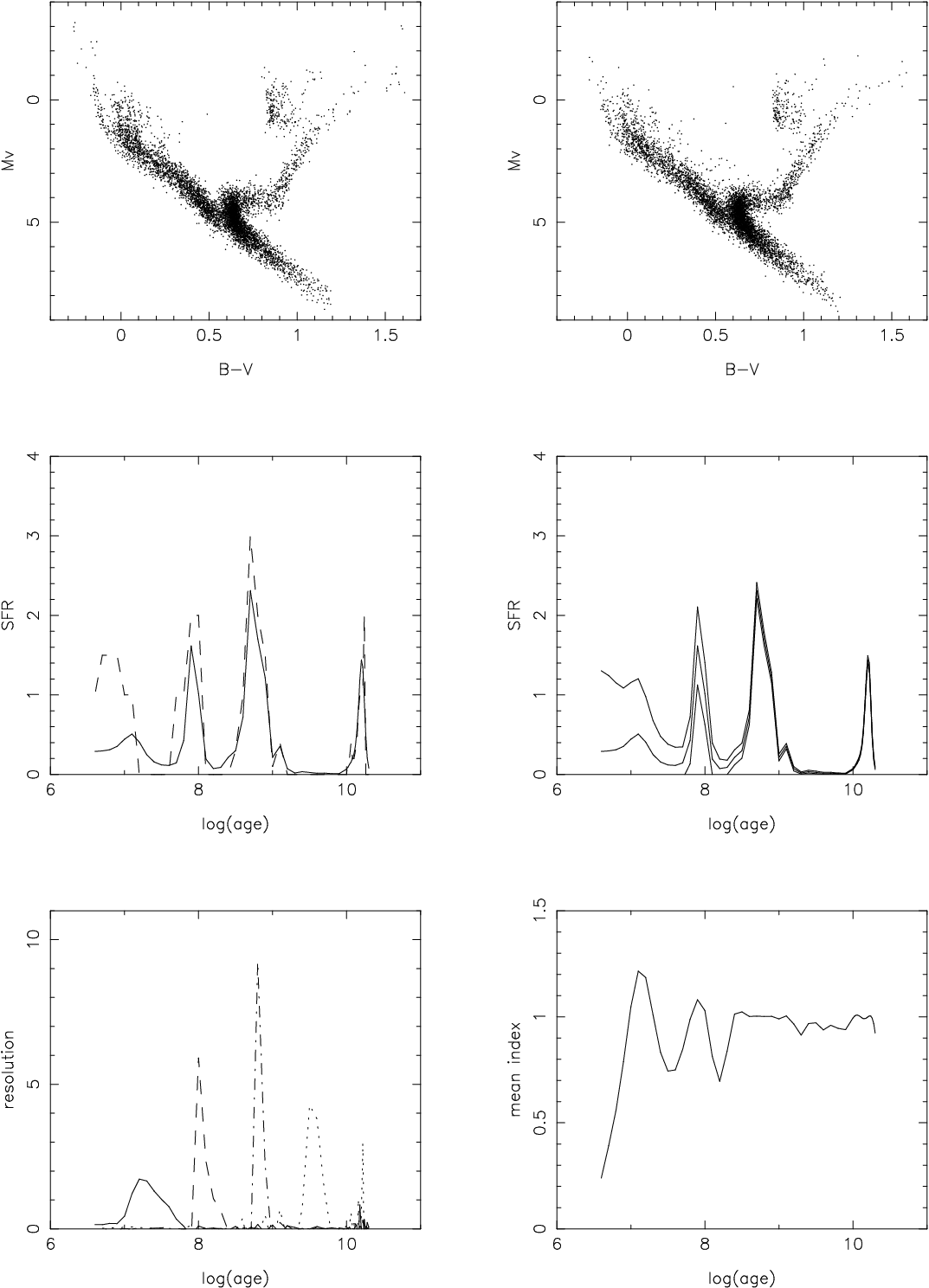}
     \caption[]{The top panels show the simulated `observed' CMD (left)
              with $\sigma_\mv=0.3$ and $\sigma_\bv=0.01$ and the
              best model (right). The middle left panel shows the
              assumed SFH (dashes) and the best model (full line),
              on the right is the best model at $\pm 1 \sigma$.
              The bottom left panel shows the resolving kernel $K(t,t')$
              as a function of age $t$ for $\log(t') = 7.3$, 8.1, 8.8, and
              9.6 (full, dashes, dot-dashes, and dots),
              and at right the mean index.
              Note that the fit for the tests is performed on the
              $B-V$ distribution only, not on the full 2-D CMD distribution.
              }
     \label{f:testa}
   \end{center}
 \end{figure*}

  As simulated observational data we took a stellar population model 
  having a fixed (solar) metallicity, a power-law IMF with slope 2.25,
  and a star formation history involving several bursts at
  different ages and with different strengths (shown as the
  full line in the middle left panel of Fig. \ref{f:testa}).

  We have to specify the parameter $\sigma_{\alpha}$ which quantifies
  how large a fluctuation we accept in the SFH. We chose it in such
  a way that one obtains reduced $\chi^2 = 0.5 ... 1.5$. Smaller values would 
  show that the data are over-interpreted, but larger values indicate 
  that the model reproduces the observations only poorly. 
  It turns out that $\sigma_{\alpha} = 1$ is a good choice. 
  The value of the smoothing length $\xi_{\alpha}$ is normalised 
  as $\sigma_{\alpha}^2 \simeq cste/\xi_{\alpha}$.

  In Fig. \ref{f:testa} we show the results obtained for the
  data to which Gaussian noise had been added with an amplitude of
  $\sigma_\mv = 0.3$~mag and $\sigma_\bv = 0.01$~mag.
  The best solution has a reduced $\chi^2=0.77$, and as
  the top panels show, the stellar population is closely reproduced.
  The slope of the IMF is $\Gamma = 2.16 \pm 0.08$,
  i.e. slightly flatter than had been assumed.
  The middle panels depict that the deduced SFH well matches all
  the features of the assumed SFH older than about 100~Myrs.
  One notes that even the small SFH feature near 1~Gyr is reproduced.
  However, the young populations are not very well represented
  because the number of corresponding `observed' stars is rather
  small ($<100$)  
  and, in addition, the $B-V$ colour of the main sequence becomes
   insensitive to age for young ages. 
  This is also evident in the bottom panels:
  while the resolving kernel shows that the relative resolution
  in age is rather good and nearly the same for ages larger than
  100~Myrs, the resolution (in $\lg t$) deteriorates for younger 
  populations. Finally, the level of information in the CMD -- as 
  measured by the mean index -- drops below about 10~Myrs.

\begin{figure*}
   \begin{center}
     \includegraphics[angle=0,totalheight=6.0in]{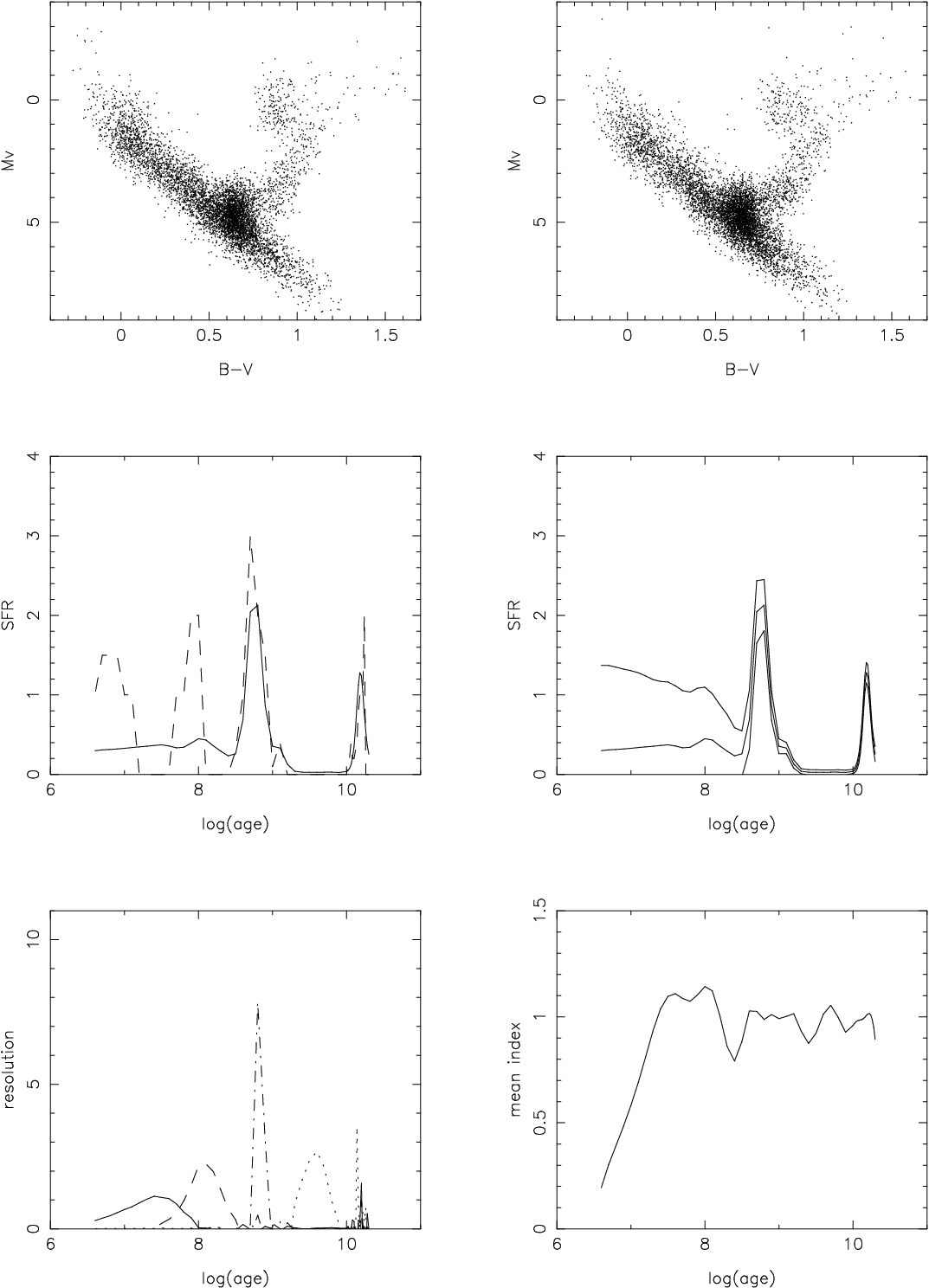}
     \caption[]{Same as Fig. \ref{f:testa}, but 
                with $\sigma_\mv=0.5$ and $\sigma_\bv=0.05$}
     \label{f:testb}
   \end{center}
 \end{figure*}

  Figure \ref{f:testb} presents the results for the same `observations',
  but degraded with larger values $\sigma_\mv = 0.5$~mag and
  $\sigma_\bv = 0.05$~mag, which is evident from the CMDs.
  The reduced $\chi^2$ obtained here is 0.64. As before, the IMF comes out 
  somewhat too flat with $\Gamma = 1.99 \pm 0.12$. The increased 
  noise level in the data destroys all the information about the 
  details of the SFH for ages up to about 1~Gyr; only the oldest 
  starbursts can be detected. The time resolution of the method 
  has become much worse than Fig. \ref{f:testa} for all ages except 
  around 1~Gyr; the drop of the mean index now occurs at about 30~Myrs.

\begin{figure*}
   \begin{center}
     \includegraphics[angle=0,totalheight=6.0in]{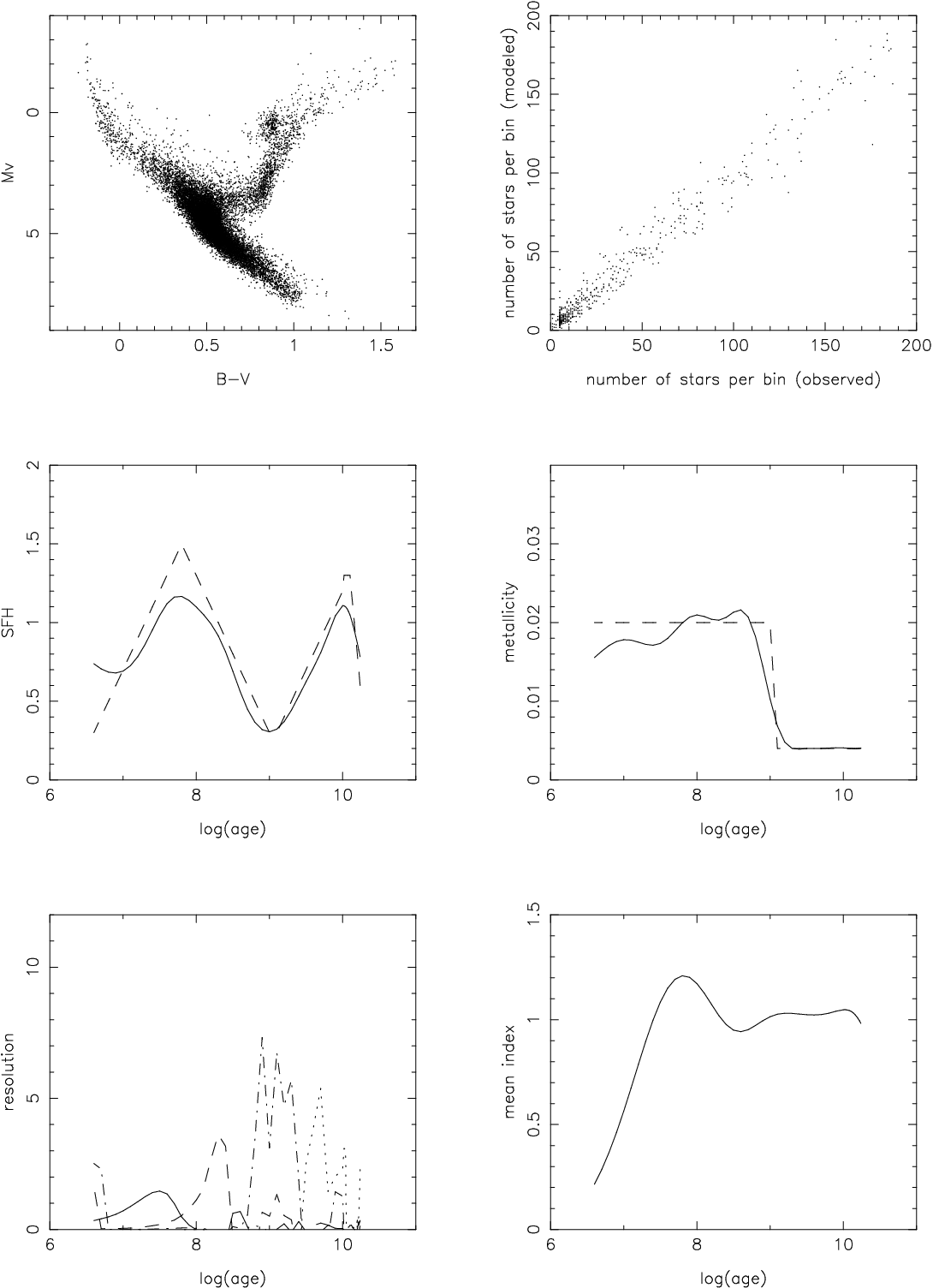}
     \caption[]{ The top panels show the simulated `observed' CMD (left)
              with $\sigma_\mv=0.3$ and $\sigma_\bv=0.01$ and the
              data fitting in number of stars per bin (right). 
              The middle left panel shows the
              assumed SFH (dashes) and the best model (full line);  
              on the right is the assumed AMR (dashes) and the best model 
              (full line).
              The bottom left panel shows the resolving kernel $K(t,t')$
              as a function of age $t$ for $\log(t') = 7.3$, 8.1, 8.8, and
              9.6 (full, dashes, dot-dashes, and dots),
              and at right the mean index}
     \label{f:testc}
   \end{center}
 \end{figure*}

  Figure \ref{f:testc} presents the results of the SFH and AMR inversion 
  from a simulated CMD composed of an old metal-poor population 
  ($Z=0.004$) and a young metal-rich one ($Z=0.02$). Because the 
  zero age main sequences are shifted with metallicity, the SFH-AMR 
  information is not degenerated in the colour-magnitude diagram. 
  The prior taken for the inverse process is $\psi_0=1$, $\alpha(t)=0$ and
  $Z(t)=0.01$, implying constant SFH and AMR. The convergence 
  occurs after about 20 iterations.

  These test results demonstrate that the method is capable of deducing 
  reliably the history of star formation in rather fine detail. It also permits 
  one to assess quantitatively the time resolution that is possible with a 
  given observational data set, as well as the information it contains.

\section{The local CMD from Hipparcos data}

   Hipparcos gives astrometry and photometry for stars within
   about 1 kpc of the Sun. Most of these stars belong to the thin
   disk, only about 5 \% of them being members of the thick disk 
   and halo (Robin et al., 1996). This sample gives us valuable
   information about the evolution of the thin disk of our Galaxy.
   This is not limited to the solar neighbourhood proper, but
   pertains also to a larger portion of the disk: because of
   the radial diffusion of stars in the disk (Wielen, 1977),
   one finds presently, near to the Sun, stars of different
   metallicities and ages.

\subsection{The Hipparcos data}

   We have selected a sample of 13520 stars from the Hipparcos
   catalogue (Perryman et al.; ESA 1997)  which meet these criteria:
   \begin{itemize}
      \item single stars
      \item the apparent magnitude limit V 
            depends on the latitude $b$ :
            $$
                 \mbox{V} \leq 7.3 +1.1 |\sin(b)| 
            $$
            for all spectral types; 
      \item observed parallaxes should be larger than 5 mas.
   \end{itemize}

   Double stars may cause some problem:
   confirmed double stars in the Hipparcos catalogue
   comprise about 20\% of all stars. Also, it is well known that
   among a stellar population about 50 \%  are double stars.
   Hence, we underestimate the presence of double stars, which could
   cause a bias in the following way: binaries will appear to be 
   located on the red side of the main sequence. They will be
   interpreted by the model as being either of higher metallicity 
   or in a more evolved state. Thus, the derived populations
   could be too old or too metal-rich. 

   The choice of the bin sizes is determined by the observational
   uncertainties of the data: For the error on the parallaxes $\pi$ we 
   take a constant value as a reasonable approximation 
   $\sigma_{\pi}=1 \mbox{(mas)}$. This results in an uncertainty of the 
   absolute magnitudes of $ \sigma_\mv = 2.17 \sigma_{\pi}/\pi$ which 
   reaches 0.5~mag at 200 pc. The error in the colour {\BV} is taken 
   as constant at 0.02 (mean error in the Hipparcos catalog).
   Therefore, we divide the HR diagram into cells of 0.5~mag in \Mv\
   and 0.02 in \BV . To prevent nearly empty cells with their
   large relative fluctuations of occupation numbers 
   to influence the results too strongly, we increase the cell size,
   if necessary, so that each bin shall contain at least 5 stars. 
   The resulting cell layout is presented in Fig. \ref{f:hrd}.

   With the limit on apparent magnitude and the local density of the
   stars, one expects only very few stars below 0.6~\Msun .
   Thus, the isochrones of Bertelli et al. (1994) which pertain only
   to stars above 0.6~\Msun\ are entirely adequate.

   Absolute magnitudes and colour indices are corrected for interstellar
   extinction with the model of Vergely (1998) which gives the opacity 
   at each point of the space in the 250~pc sphere surrounding the Sun, 
   derived with a tomographic technique from Hipparcos data and 
   Str\"omgren photometry.

   \begin{figure}
     \begin{center}
      \includegraphics[angle=0,totalheight=4.0in]{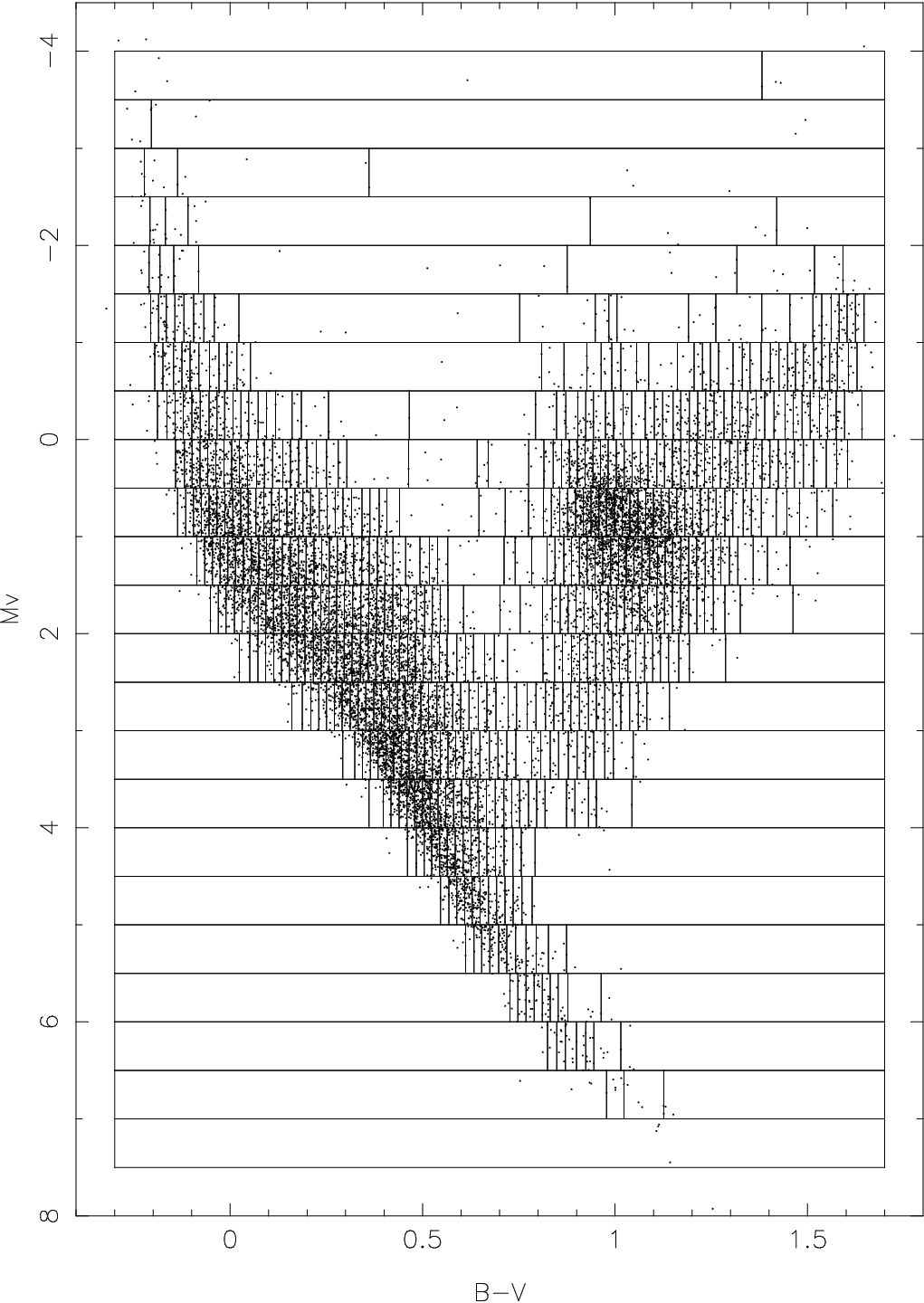}
       \caption{The colour-magnitude diagram of the single stars of the
                Hipparcos catalogue, with the adopted binning in both
                dimensions}
      \label{f:hrd}
     \end{center}
   \end{figure}

\subsection{The base models $F(u)$}

   The construction of the base models follows that described
   in Sect. \ref{s:base}, but we now take into account the 
   above-mentioned selection criteria for the observational data.

   Furthermore, for solar neighbourhood stars, we must also take 
   into account that older stars are distributed up to greater heights 
   above the Galactic plane than younger ones, because the amplitude 
   of their oscillatory motions perpendicular to the plane increases 
   with age (cf. Wielen 1977). 

   In this study, we restrict our SFH-AMR analysis to the
   thin disk stars. The contamination by thick disk population
   stars can be neglected in a first approach, because only
   about 5\% of stars belong to the thick disk in the Hipparcos
   sample. The age-dispersion relation given by Gomez et al. (1997)
   is valid only for thin disk population stars and shows clearly
   that there is no significant dynamic evolution of the
   thin disk, after 4-5 Gyr. More accurately, the age-dispersion relation
   is given for old thin disk stars with age $<$ 9 Gyr and saturated at
   15-17 km/s.

   The relation between age $t$ ([Gyr]) and
   dispersion of the vertical velocity component $W$ ([km/s]) has
   recently been derived from Hipparcos data (G\'omez et al., 1997):
   $$
        \sigma_W(t) = 17.0-12.0  \exp(-{t/2})
   $$ 
   Assuming isothermality,  the density of coeval stars  of age $t$
   is constrained by  the vertical potential $\phi(z)$ at  height $z$ as:
   $$
          \rho(t,z) = \rho(t,0)
              \exp\left(-{\phi(z) \over \sigma_W^2(t)}\right)
   $$
   where $\rho(t,0)$ is the density at $z=0$ and $\phi(z)$ is the potential. 

   Recent determinations of the  total local mass density $\rho_0$, based
   on  Hipparcos  data,  give  $\rho_0$  from  $0.076~  M_\odot  pc^{-3}$
   (Cr\'ez\'e et  al, 1997,1998) to $0.10~ M_\odot  pc^{-3}$ (Holmberg \&
   Flynn,   2000).   The   surface  mass   density  is   determined  from
   measurements of the force at larger  distances (0.5 to 1 kpc) from the
   galactic plane: recent determinations of $\Sigma_0$ range from $48$ to
   $56~ M_\odot pc^{-2}$, (Kuijken \& Gilmore 1989, Flynn \& Fuchs 1994).
   These two independent sets of  constraints on the vertical potential may
   be combined in a single formula:
   \begin{equation}
       \label{e:potential}
          \Phi(\rm  z)= 0.027~ \Sigma_0  (\sqrt{\rm  z^2+ \rm  D^2}  - 
                    \rm D  + \rho_{\rm eff}~ z^2)
   \end{equation}
   Here, we  will use the coefficients $D=240~\rm  pc$, 
   $\Sigma_0=48 M_\odot pc^{-2}$ and
   $\rho_{\rm eff}=~0.0105~  M_\odot pc^{-3}$, this implies  a local mass
   density of $0.081~ M_\odot pc^{-3}$  and a disc surface mass density
   of $48~ M_\odot pc^{-2}$.
   
   \begin{figure}
     \begin{center}
      \includegraphics[angle=0,totalheight=4.0in]{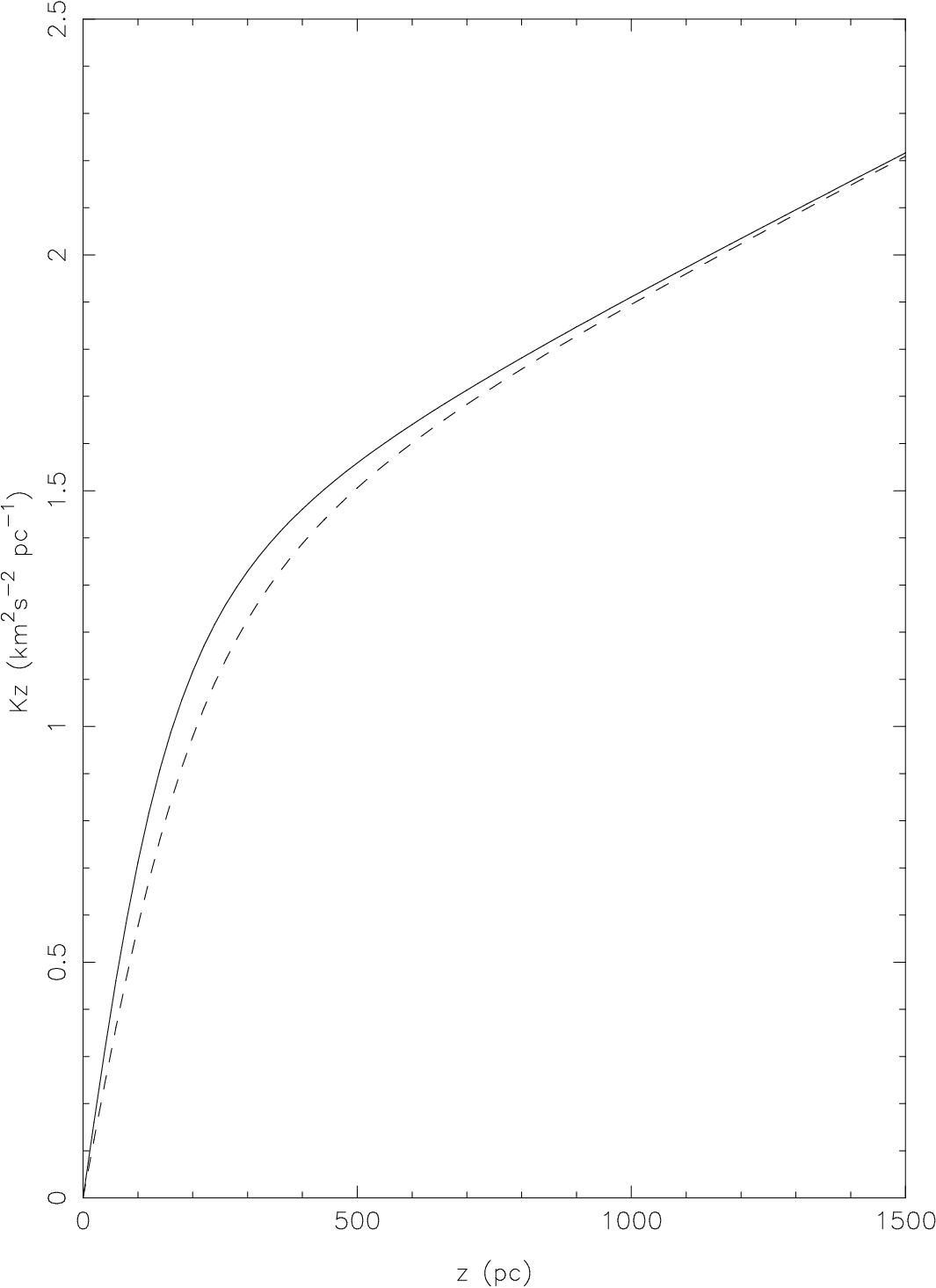}
       \caption{The vertical potential (continuous line) force 
                    given by Kuijken \& Gilmore (1989) and the modified force 
                    (dashed line)  used here in order to include the most recent 
                    determinations of a local mass density in the solar 
                    neighbourhood.}
      \label{f:hrd}
     \end{center}
   \end{figure}

\subsection{Determination of IMF slope and SFH}

   We first consider the metallicity to be constant, at solar value 
   $Z = 0.02$. The resultant fit is not good (reduced $\chi^2 = 4.9$), so
   that  the model  is  to be  rejected.

   The derived  IMF slope  is $\Gamma = 3.2 \pm 0.1$, steeper than 
   Salpeter's, but quite close to the slope  of 2.7 found by  e.g. Kroupa, 
   Tout \&  Gilmore (1993) in the  solar neighbourhood for stars above 
   1 \Msun  .  It  is these stars that contribute to most of the data in 
   our sample.

\begin{figure}
    \begin{center}
      \includegraphics[angle=0,totalheight=5.0in]{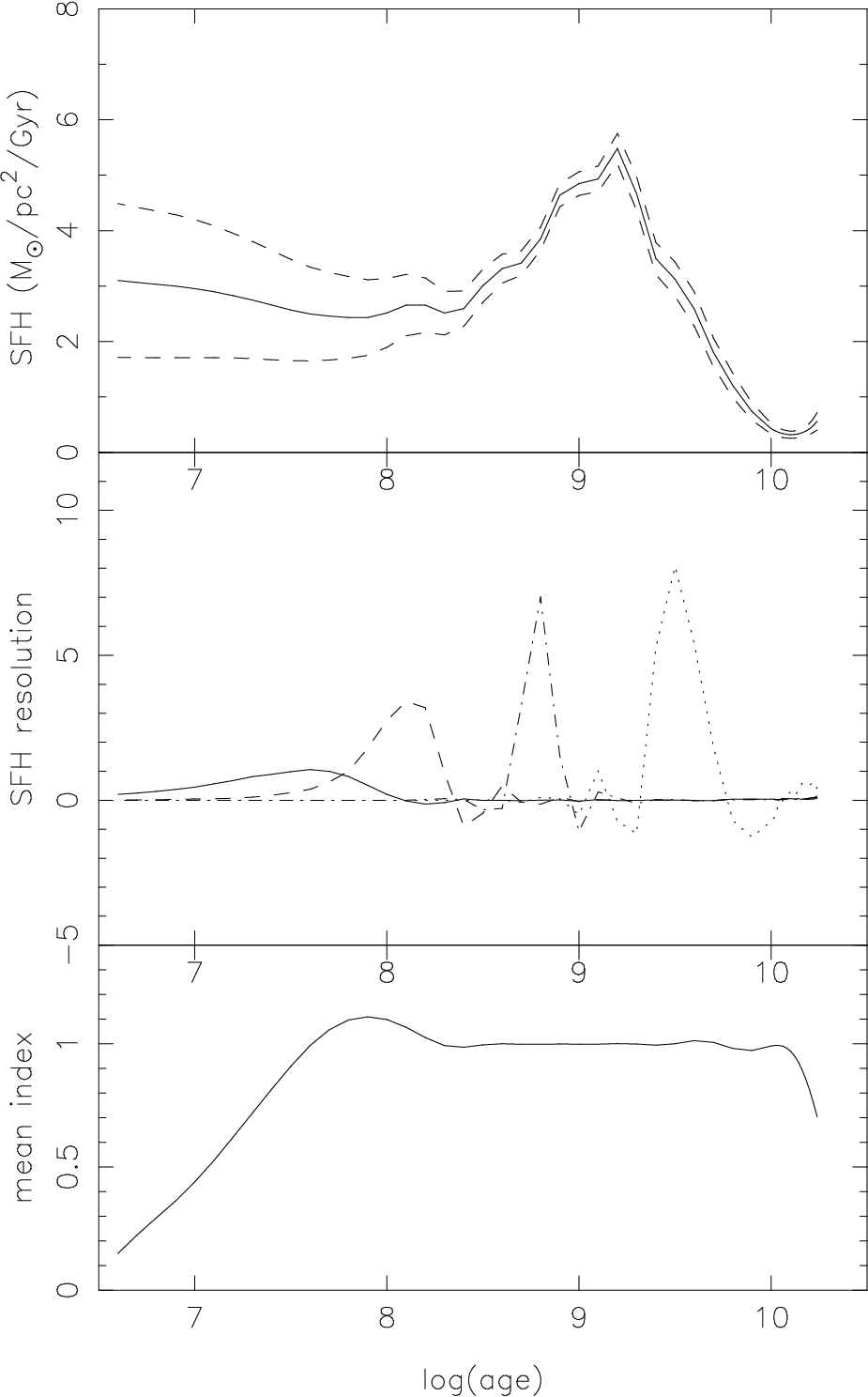}
      \caption[]{The star formation history of the solar neighbourhood,
                 assuming a constant stellar metallicity $Z = \zsun$.
                 From top to bottom: The derived SFH with standard
                 errors, the resolving kernel $K(t,t')$
                 as a function of age $t$ for $\log(t') = 7.3$, 8.1, 8.8, and
                 9.6 (full, dashes, dot-dashes, and dots)
                 and the mean index.}
       \label{f:solara}
    \end{center}
  \end{figure}

   Main sequence magnitudes lie between about 6 and $-2$ mag,
   corresponding to masses between 0.6 and 8 \Msun ,
   and lifetimes of greater than 40 Myrs. Thus, information about
   the last 40 Myrs is not present. This is seen in the
   turn-down of the mean index (cf. Fig. \ref{f:solara}).
   The prior taken for the inverse process is (see Eqn. \ref{par1}) 
   $\alpha(t)=0$, implying a constant SFH. 

   The deduced SFH is presented in Fig.~\ref{f:solara}.
   The resolution is rather poor; from the widths of
   the kernels one gets 0.4~dex in the relative age. The a posteriori 
   error is quite large for the youngest populations and the results are 
   not significant for ages below 30 Myrs, as clearly shown by the
   low mean index which drops at about 50 Myrs.

\subsection{Simultaneous determination of SFH and AMR}

   As the model with constant metallicity has to be rejected, 
   we now investigate models with a fixed IMF slope of $\Gamma = 3.0$,
   but leaving both the SFH and AMR as free parameters.   
   As priors for the inverse process we take $\alpha(t)=0$ and  
   $Z(t)=0.02$, i.e. constant SFH and AMR. 

   If we assign equal weights to main sequence and giant stars,
   we obtain a rather poor value of reduced $\chi^2 = 3.7$. The model population
   does not fit the main sequence, in particular below it, many metal-poor
   stars are predicted. The reason is that the prominent red clump --
   composed of horizontal branch stars of nearly solar metallicity --
   contains many relatively blue stars which in the context of the
   presently-used stellar tracks correspond to low metallicity isochrones. 

\begin{figure}
    \begin{center}
     \includegraphics[angle=0,totalheight=4.0in]{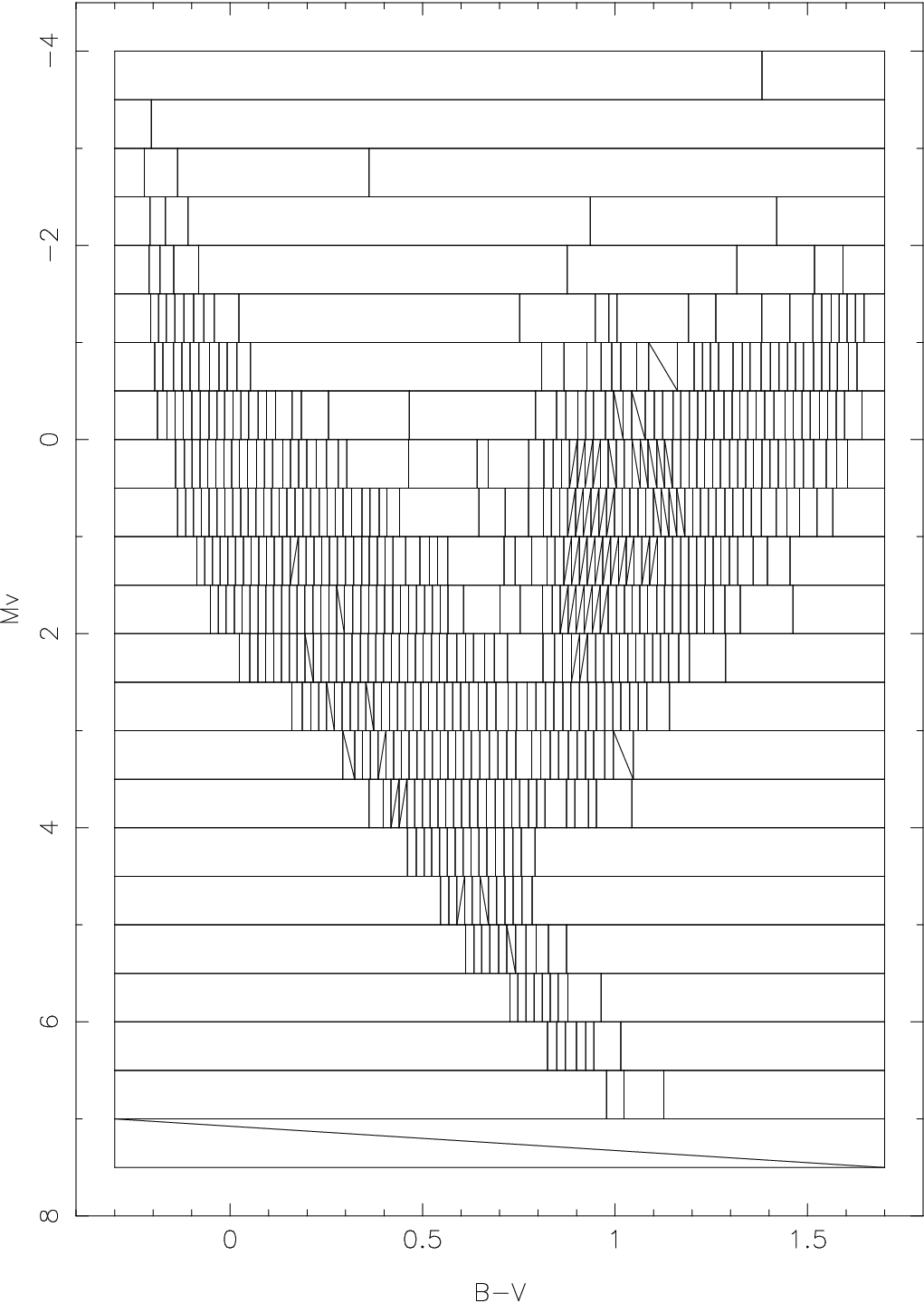}
      \caption[]{Comparison of model and observed CMD: 
                 The /  ($\backslash$) indicate the bins whose number 
                 of model stars is smaller (greater)
                 by at least $3\sigma$ than the observations}
      \label{f:fit}
    \end{center}
  \end{figure}

   If we reduce the weight for the giants by a factor of two, a rather
   satisfactory fit with reduced $\chi^2 = 1.9$ is obtained. Figure \ref{f:fit}
   shows that the main sequence is well reproduced except for the
   odd bin, but that a major discrepancy remains with the red clump 
   where the observed stars tend to be systematically bluer than the
   model. Very recently, Girardi et al. (2000) published updated isochrones.
   Detailed comparison with the isochrones of Bertelli et al. (1994)
   reveals that the horizontal branch is shifted to the blue by about
   0.05 mag in \BV\  which is the right amount to remove the discrepancy
   (as shown by test calculations).

   However, this shift could be explained by a dispersion in the AMR 
   (not taken into account in this study) or the presence of unresolved 
   giant binary stars.

  \begin{figure}
    \begin{center}
      \includegraphics[angle=0,totalheight=5.0in]{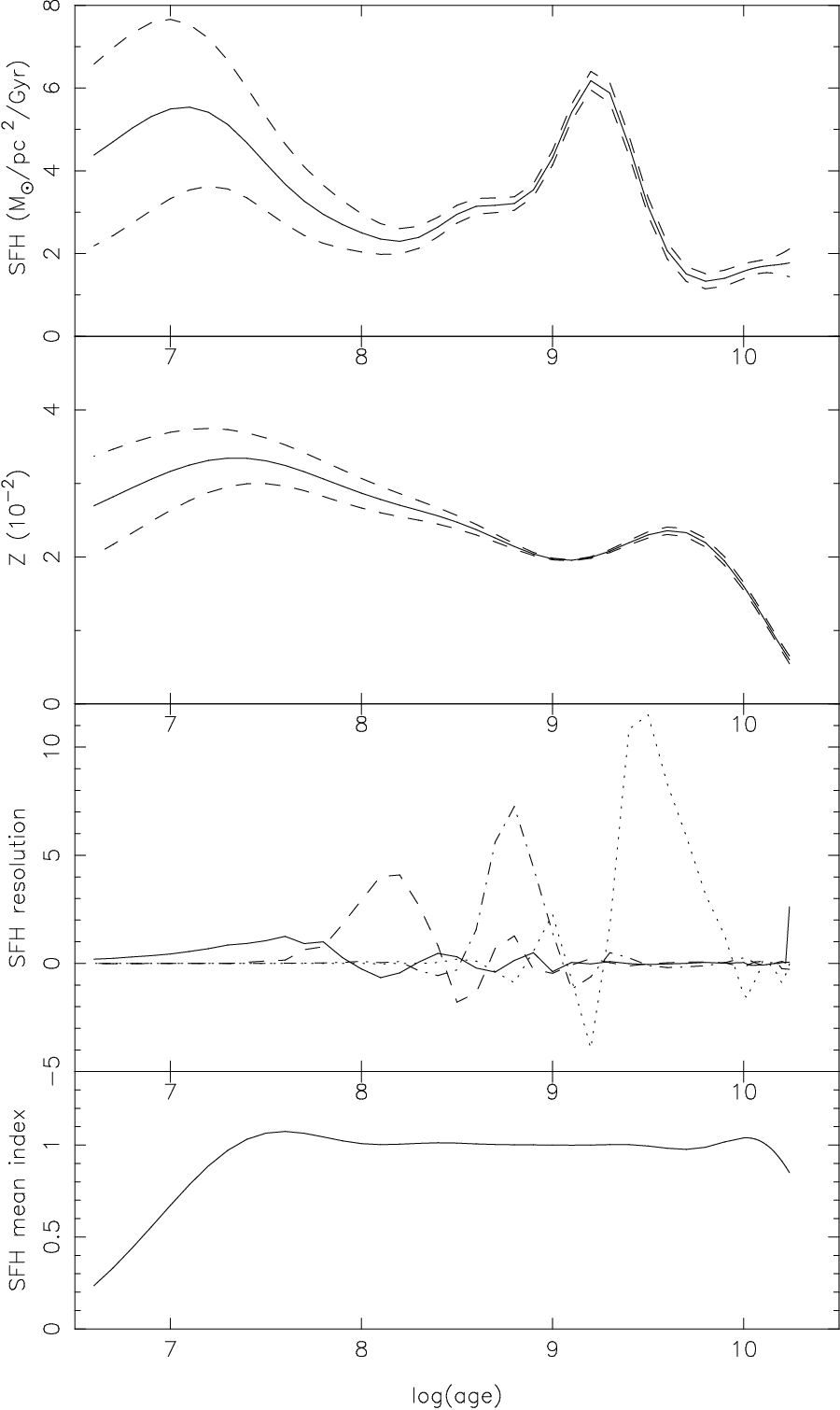}
      \caption[]{Similar to Fig.\ref{f:solarb}, but for the simultaneous
                 inversion for SFH and AMR. The second panel from the top
                 shows the deduced AMR with standard errors}
      \label{f:solarb}
      \end{center}
  \end{figure}

\subsubsection{The star formation history}

  The SFHs obtained from the two approaches are quite similar,
  having a rather broad peak near an age of 1.6 Gyr.
  The SFH at the present time deduced with the constant metallicity
  model is rather low, while in the model with variable metallicity
  (Fig. \ref{f:solarb}) it is almost as high as this peak. Note
  howeverthat the information available allows a reliable deduction
  of the SFH with good resolution only for ages larger than
  about 100~Myrs. The age resolution achieved in the second model 
  is somewhat poorer, due to the information now being used to 
  derive the AMR as well.

  A first preliminary analysis of the Hipparcos CMD by Bertelli
  et al. (1997) yielded that star formation occurs between ages
  of 0.1 and 10~Gyrs, with indication of a discontinuity at
  about 1.5~Gyr. Also, the broad red clump indicated a spread in
  metallicity. From a rather qualitative comparison they find that
  the SFH should have been constant or increasing from 10 to
  1.5~Gyrs, and be reduced by a factor 2 or 3 after that time.
  Our results fully confirm these findings, showing a broad and
  rather well-defined peak of star formation activity at about
  1.5~Gyrs with the SFH being substantially lower before and after. 

  Bertelli and Nasi (2000) use a comparable set of Hipparcos stars to
  analyse the SFH in the solar neighborhood. Their principal conclusion
  is that the SFH increases, in a broad sense, from the beginning to the
  present time; that is in agreement with the SFH obtained in this
  paper. They have tested different IMF slopes and preferred the IMF Salpeter
  slope in order to fit the main sequence. However they underline that
  an IMF slope of 3.35 is an acceptable solution, particulary to find
  the best ratio between the number of stars in the He burning phase
  and that in the main sequence phase. Our analysis shows that this IMF 
  slope can fit very well the main sequence, too.
 
  Hernandez et al. (2000) use an inverse method to derive the SFH
  over the last 3 Gyr. They show clearly a bump in the SFH at about
  2 Gyr. Similar to their results we do not find the decreasing SFH activity
  between 1 and 2 Gyr as found by Rocha-Pinto (1999) from 
  chromospheric activity. 

  The determinations of the IMF (cf. Scalo 1986) yield that the
  present SFR should be nearly equal to the past average SFR.
  Our study gives the same result:
  $$
     {\int_0^{10 {\rm Gyr}} \psi(t) dt / 10{\rm Gyr} \over \psi (0)}=0.9  .
  $$

  A quite different approach can be obtained from the chemical
  evolution of the solar neighbourhood:
  Rocha-Pinto \& Maciel (1997) derived from the metallicity
  distribution of the G-dwarfs and the AMR the SFH
  (cf. Eqn. \ref{e:gdwarf} below) taking into account the
  observational scatter. They demonstrate that for several observed
  AMRs one arrives at rather similar SFHs: a broad peak at about 8~Gyrs
  ago with about twice the present SFH and an epoch of low SFH (one half
  or less of the present value) between 1 and 3~Gyrs ago. 
  As this approach requires the {\it slope} of the already rather uncertain
  and noisy AMR, the results are quite sensitive to the adopted AMR,
  e.g. the mean AMR of Edvardsson et al. (1993) gives a
  nearly flat SFH. At an age of 1.5~Gyrs, when the Hipparcos
  CMD gives a peak SFH, Rocha-Pinto \& Maciel rather find a minimum.
  Though their result seems statistically significant, one should
  emphasize that this approach can yield only a resolution which is
  nearly constant in (linear) time, because the metallicity changes
  nearly linearly in time with some flattening near the present.
  On the contrary, the distribution of stars in a CMD is sensitive to
  the logarithm of the age. Therefore we feel that the Hipparcos CMD
  gives a stronger constraint on the SFH in the past few Gyrs. Certainly,
  this discrepancy between chemical evolution and stellar population
  needs further attention.

\begin{figure}
     \begin{center}
       \includegraphics[angle=0,totalheight=4.0in]{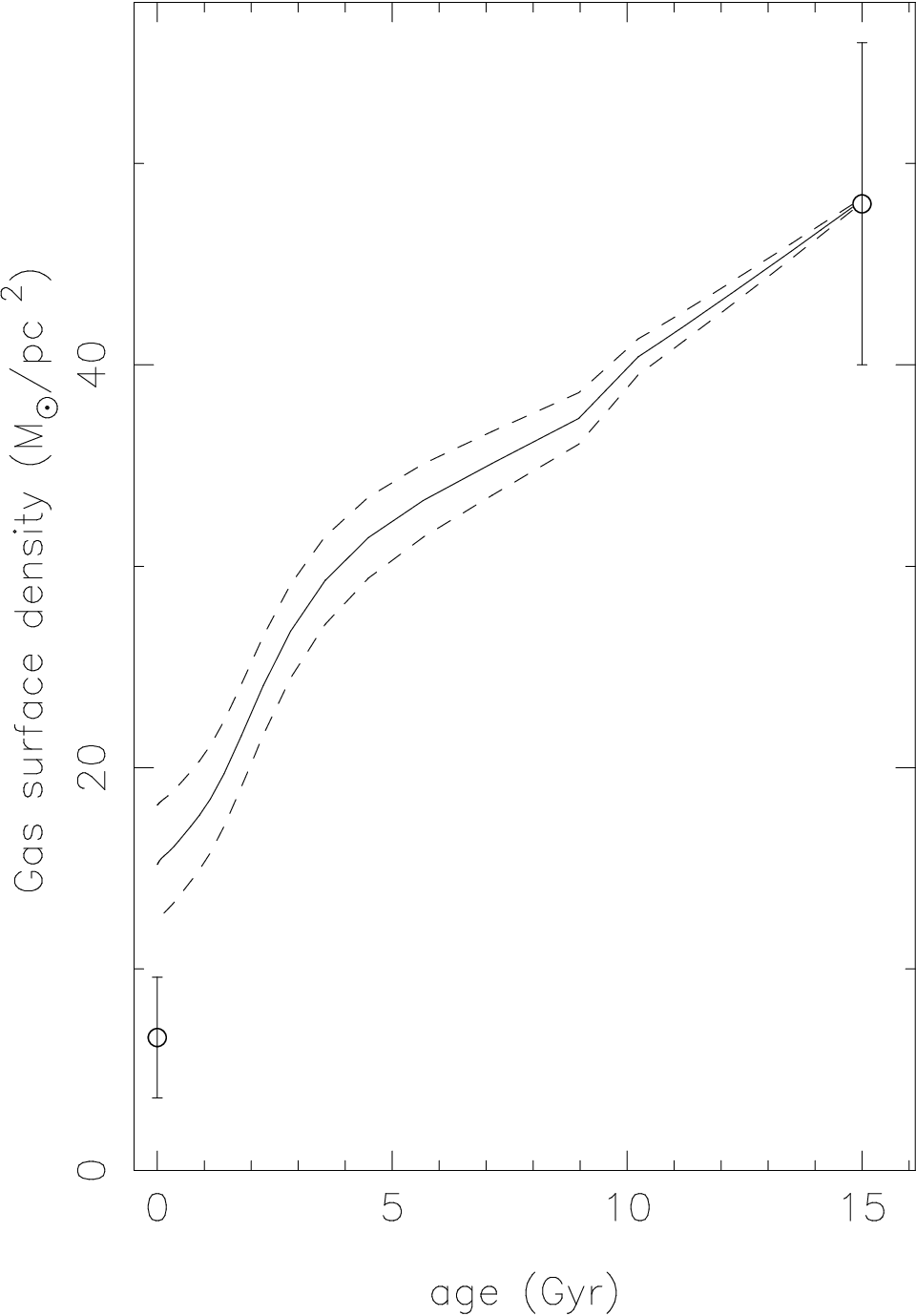}
       \caption[]{Evolution of the gas surface density in the solar
                  neighbourhood, computed with our SFH (solid line).
                  The two dotted lines indicate the error at the
                  $\pm 1 \sigma$ level. The circle at 0~Gyr is the present 
                  surface gas density (Rana, 1992). The circle at 15~Gyr 
                  is the present dynamical density (Kuijken \& Gilmore, 1991). 
                  For both values 1 $\sigma$ error bars are given}
       \label{f:gas}
      \end{center}
    \end{figure}

  As a further check we investigate whether our SFH is in agreement
  with the past gas consumption in the solar neighbourhood:
  in a closed-box model the evolution of the surface gas density  
  $\Sigma_g$ is given by (Tinsley, 1980) :
  \begin{equation}
       \label{e:gasfraction}
              \frac{d\Sigma_g}{dt}=-(1-R) \psi(t)  .
  \end{equation}
  The constant $(1-R)$ is the locked-up mass fraction, which is 
  about 0.8 for a Salpeter IMF, and rather insensitive to the IMF. 
  The total surface density of the mass present in the disk was 
  derived from stellar dynamics by Kuijken \& Gilmore (1991) 
  as 48 $M_{\odot}{\rm pc}^{-2}$. The current gas surface density 
  is 6.6 $M_{\odot}{\rm pc}^{-2}$ estimated by Rana (1992) as
  the sum of molecular and atomic hydrogen. Wyse \& Gilmore (1995) 
  argue for a somewhat higher present day gas fraction of 25 percent.
  Using the total mass as the initial condition, we integrate
  Eqn. \ref{e:gasfraction}. The evolution of the gas density shown 
  in Fig. \ref{f:gas} matches the present day gas fraction rather well.

\subsubsection{The age-metallicity relation}

  In their preliminary study, Bertelli et al. (1997) emphasized that
  the colour extension of the horizontal branch clump as well as the
  width of the main sequence provide evidence for an extended range 
  ($Z = 0.008 ... 0.03$) of metallicities in the stellar population.
  This is corroborated by our study, in the sense that a much better fit is
  obtained, if one allows for a variation of the metallicity.

\begin{figure}
     \begin{center}
       \includegraphics[angle=0,totalheight=4.0in]{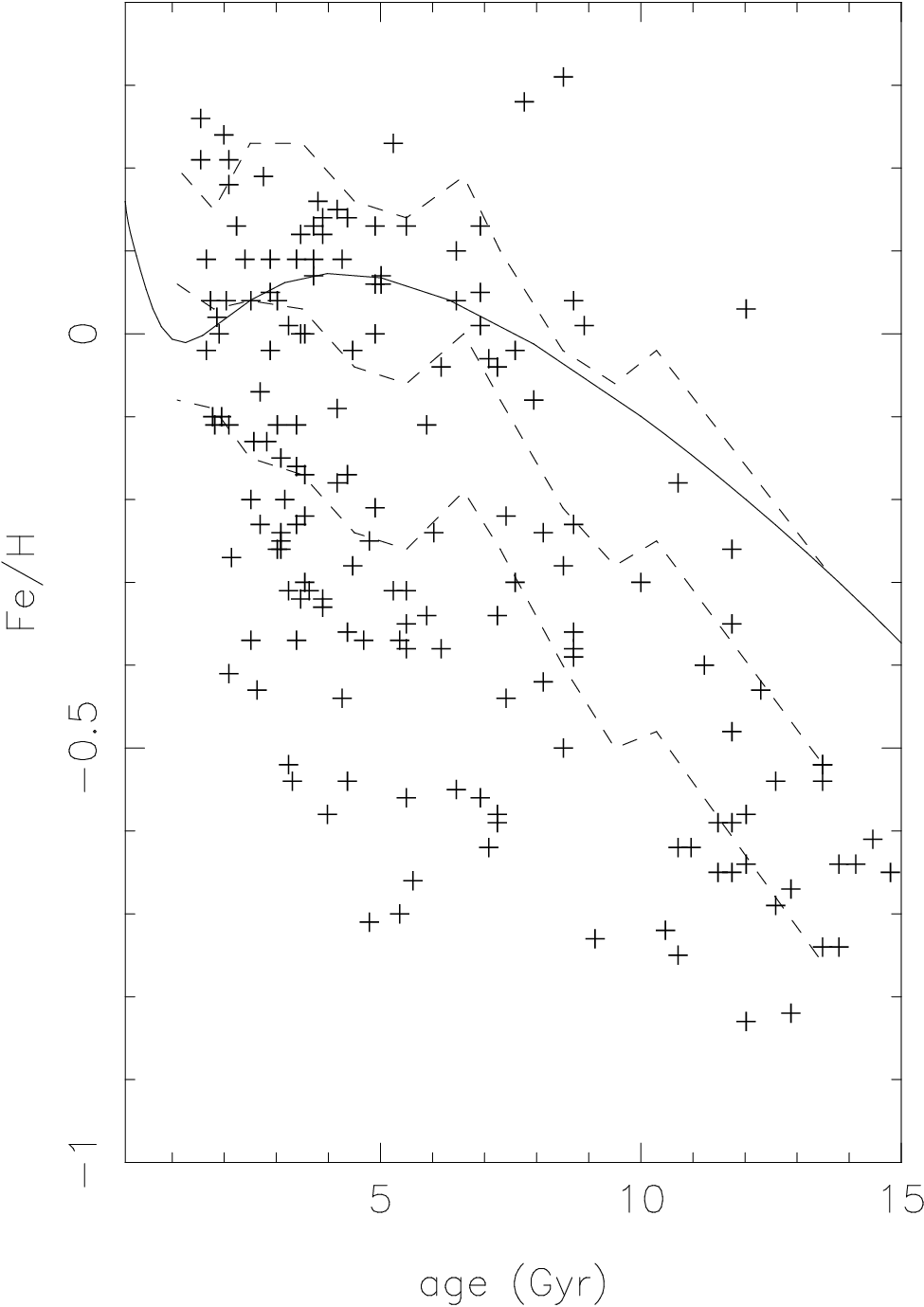}
        \caption[]{Age-metallicity relations obtained with the inversion
                  (full line) in comparison with the relation of
                  Meusinger et al. (1991) -- dashed lines giving
                  the average and $\pm 1 \sigma$ curves --
                  Crosses are individual stars from
                  Edvardsson et al. (1993).}
       \label{f:amr}
      \end{center}
    \end{figure}

  The derived relation between age and metallicity is compared 
  in Fig. \ref{f:amr} with those obtained by others. The metallicity
  increases with time, reaching a peak (at solar metallicity) 
  at an age of 4 Gyr; thereafter it descends somewhat, going through
  a minimum at 1~Gyr, and then rising again. 
  Though Fig. \ref{f:solarb}
  shows that this wavy structure is significant, we prefer not to place
  too strong an emphasis on it, because on one hand the information is 
  used for both SFH and AMR, that are treated as independent functions. 
  With this greater amount of freedom, a good fit may well be achieved.
  On the other hand, the finding that the peak SFH 
  at an age of 1.5~Gyr is not associated
  with or closely followed by a strong increase in the metallicity,
  as would be expected from chemical evolution, points well in
  the same direction. 
  Also, the time resolution in the SFH is rather coarse
  (perhaps 0.3 dex) which also pertains to the AMR.
  
  The causal link betwen star formation and metal enrichment
  could and should be used as a constraint to make the SFH and AMR 
  consistent with each other. While it is conceivable that it be incorporated
  in our method, it is not the aim of the present paper to do that.

\begin{figure}
     \begin{center}
       \includegraphics[angle=0,totalheight=4.0in]{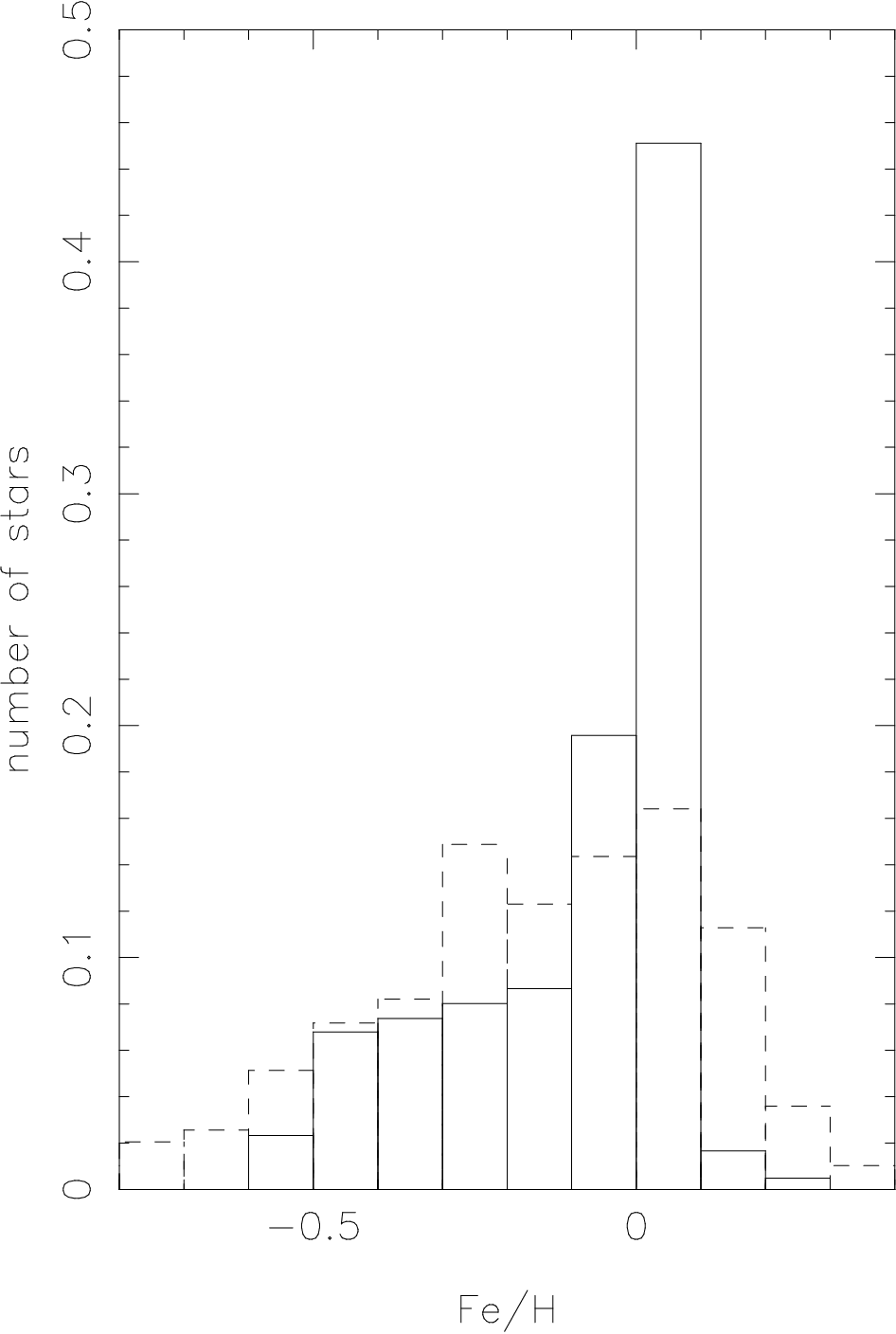}
       \caption[]{Metallicity distribution function for the G dwarfs  
                  (Rocha-Pinto \& Maciel, 1997, dashed line) and 
                  computed from our SFH and AMR (full line)} 
       \label{f:gdwarfs}
      \end{center}
    \end{figure}

   The histogram of the metallicities of the G dwarfs is a powerful 
   constraint for the chemical evolution of the disk (cf. Pagel, 1997). 
   From our SFH and the time derivative of our AMR we compute this 
   distribution function as
   \begin{equation}
       \label{e:gdwarf}
       d N_{*} / d \lg Z = {d N_*/dt \over d\lg Z/dt}
            \propto Z \cdot \psi(t(Z))/(dZ/dt)
   \end{equation}
   which is shown in Fig. \ref{f:gdwarfs} in comparison to the
   histogram from Rocha-Pinto \& Maciel (1997).
   Note that the distribution from Rocha-Pinto \& Maciel suffers
   from the additional dispersion inherent to the measurements, that we
   did not try to simulate.
   Our distribution has a sharp peak near solar metallicity, because
   our AMR is rather flat at an age around 1 Gyr.
   This could be due to errors in the metallicities as 
   determined with Str\"omgren photometry.
   Using their sample of G dwarfs,  
   Rana \& Basu (1990) show that the intrinsic dispersion of metallicity 
   reaches $0.24 \pm 0.10$~dex at a given age. 

\section{Conclusions}
    The method presented here permits us to extract SFH, IMF slope and 
    age-metallicity relations in a coherent way taking into account 
    the selection function in apparent magnitude and the dispersion 
    velocity of the stars. Indicators like the mean index and 
    the resolution give a good idea of the quality of the results. It shows  
    clearly that the mixed populations cannot be separated with 
    an infinite accuracy and that the results depends on the 
    accuracy of the evolutionary tracks. 

  The observed Hipparcos HR diagram is well reproduced by the model 
  at the main sequence level while the horizontal giant branch is not 
  correctly fit. This could be due to the poor knowledge of the parameters that 
    constrain the evolution of old stars. Also, our results could be biased by 
    the presence of undetected binary stars or peculiar colour behaviour of 
    A stars in the main sequence.  

    The derived SFH shows enhanced star formation about 1 to 2 Gyrs ago,
    while the present SFR is about equal to the past average SFR. This agrees
    well with the overall gas consumption in the solar neighbourhood.
    The AMR displays a systematic rise in metallicity with time. Both the
    AMR and SFH are fairly consistent with the paucity of metal-poor
    G dwarfs.

    The principal result found in this study is that the thin 
    disk of the Milky Way is rather young and has a solar metallicity. 

\begin{acknowledgements}
     We thank G.Lapierre for stimulating discussions that triggered 
     this study. This work is part of the thesis of JLV who acknowledges
     support of a fellowship from {\it Minist\`ere de l'Education
     Nationale} (France).
\end{acknowledgements}


\begin{thebibliography}{99}

\bibitem{} Backus G., Gilbert F.: 1970, 
             Phil. Trans. R. Soc. London 266, 123
\bibitem{} Bertelli G., Bressan A., Chiosi C., Fagotto F., 
             Nasi E.: 1994, A\&AS 106, 275
\bibitem{} Bertelli G., Nasi E., Bressan A., Chiosi C.: 
             1997, ESA SP-402, 501
\bibitem{} Bertelli G. and Nasi E.:
             2000, ApJ in press (astro-ph/0011126) 
\bibitem{} Craig I.J.D., Brown J.C.: 1986,
             {\it Inverse Problems in Astronomy},
             Adam Hilger Ltd., Bristol and Boston
\bibitem{} Cr\'ez\'e M., Chereul E., Bienaym\'e O., Pichon C.: 
           1998, A\&A 329, 920
\bibitem{} Dolphin A.: 1997, New Astronomy 2, 397
\bibitem{} Dolphin A.: 2001, MNRAS in press (astro-ph/0112331)
\bibitem{} Edvardsson B., Andersen J., Gustafsson B., 
             Lambert D.L., Nissen P.E., Tomkin J.: 1993, A\&A 275, 101
\bibitem{} ESA: 1997, {\it The Hipparcos and Tycho Catalogues, 
             Volume 1}, SP-1200, 131
\bibitem{} Flynn C., Fuchs B.: 1994, MNRAS 270, 471
\bibitem{} Gallart C., Aparicio A., Bertelli G., Chiosi C.:
             1996, AJ 112, 1950
\bibitem{} Girardi L., Bressan A., Bertelli G., Chiosi C.: 2000, 
               A\&AS 141, 371 
\bibitem{} G\'omez A.E., Grenier S., Udry S., Haywood M., 
             Meillon L., Sabas V., Sellier A., Morin D.:
             1997, ESA SP-402, 621
\bibitem{} Greggio L., Marconi G., Tosi M., Forcadi P.: 1993, AJ 105, 894 
\bibitem{} Harris J., Zaritsky D.: 2001, ApJS 136, 25
\bibitem{} Haywood M., Robin A., Cr\'ez\'e M.: 1997, A\&A 320, 428 
\bibitem{} Hernandez X., Valls-Gabaud D., Gilmore G.: 1999, MNRAS 304, 705
\bibitem{} Hernandez X., Valls-Gabaud D., Gilmore G.: 2000, MNRAS 316, 605
\bibitem{} Holmberg J., Flynn C.: 2000, MNRAS 313, 209
\bibitem{} K\"oppen J.: 1994, A\&A 281, 26
\bibitem{} Kroupa P., Tout C.A., Gilmore G.: 1993, MNRAS 262, 545
\bibitem{} Kuijken K., Gilmore G.: 1991, ApJ 367, L9
\bibitem{} Kuijken K., Gilmore G.: 1989, MNRAS, 239, 605 
\bibitem{} Meusinger H., Reimann H.G., Stecklum B.: 1991, A\&A 245, 57 
\bibitem{} Meyer D.M., Jura M.J., Hawkins I., Cardelli J.A.: 
             1994, ApJ 437, L59   
\bibitem{} Ng Y.K., Bertelli G.: 1998, A\&A 329, 943 
\bibitem{} Perryman M.A.C. et al.: 1997, A\&A 323, L49
\bibitem{} Pagel B.E.J.: 1997, "Nucleosynthesis and Chemical Evolution 
            of Galaxies", Cambridge University Press
\bibitem{} Rana N.C., Basu S.: 1990, Ap\&SS 168, 317
\bibitem{} Robin A.C., Haywood M., Cr\'ez\'e M., Ojha D. K., 
             Bienaym\'e O.: 1996, A\&A 305, 125 
\bibitem{} Rocha-Pinto H.J., Maciel W.J.: 1997,  MNRAS 289, 882
\bibitem{} Rocha-Pinto H.J., Scalo J., Maciel W.J., Flynn C.: 2000a,
             ApJ 531, L115
\bibitem{} Rocha-Pinto H.J., Maciel W.J., Scalo J., Flynn C.: 2000b,
             A\&A (astro-ph/0001382)
\bibitem{} Salpeter E.E.: 1955, ApJ 121, 161
\bibitem{} Saporta G.: 1990, {\it Probabilit\'es}, Editions Technip, Paris
\bibitem{} Scalo J.M.: 1986, Fund. Cosmic Phys. 11, 1
\bibitem{} Tarantola A., Valette B.: 1982a,  J.Geophys. 50, 159
\bibitem{} Tarantola A., Valette B.: 1982b,
             Rev. Geophys.\& Space Phys. 20, 219
\bibitem{} Tinsley B.M. : 1980, Fund. Cosmic Phys. 5, 287
\bibitem{} Tolstoy E., Saha A.: 1996, ApJ 462, 672
\bibitem{} Twomey S.: 1977, "Introduction to the Mathematics of Inversion 
                in Remote Sensing and Indirect Measurements", Dover, New York 
\bibitem{} Vergely J.-L.: 1998, Ph.D. Thesis, Obs. de Strasbourg 
\bibitem{} Wielen R.: 1977, A\&A 60, 263
\bibitem{} Wyse R.F.G, Gilmore G., 1995, AJ 110, 2771

\end{thebibliography}
\end{document}